\documentclass[11pt,a4paper,english,superscriptaddress,aps,prd,preprint]{revtex4-1}
\usepackage{amssymb}
\usepackage{amsmath}
\usepackage{graphicx}
\usepackage{babel}
\usepackage{hyperref}
\usepackage{color}

\begin{document}

\title{Lorentz-violating quantum electrodynamics in two-dimensional aether-superspace}

\author{A.~C.~Lehum}
\email{andrelehum@ect.ufrn.br}
\affiliation{Escola de Ci\^encias e Tecnologia, Universidade Federal
  do Rio Grande  do Norte\\
Caixa Postal 1524, 59072-970, Natal, Rio Grande do Norte, Brazil}

\begin{abstract}
The two-dimensional aether-superspace is constructed and the superfield techniques are applied to the study of dynamical generation of mass in the Lorentz-violating supersymmetric quantum electrodynamics in two dimensions of spacetime. It is shown that such model presents a dynamical generation of mass to the gauge aether-superfield and its dispersion relation has the structure similar of the CPT-even Lorentz-breaking models. 
\end{abstract}

%\pacs{11.30.Pb, 11.30.Cp}

\maketitle

\section{Introduction}\label{intro}

The Standard Model (SM) of particle physics is very successful in describing matter and its interactions. Despite its success, there are fundamental questions that the SM does not adequately explain, such as the origin of neutrino masses, why gravity is so weak if compared with the other forces, what is dark matter and dark energy, strong CP problem, etc... These unexplained phenomena have motivated physicists to search for extensions of the SM, where the most studied ones are the String Theory, Supersymmetry (SUSY), Extra-dimensions and Lorentz violations of the SM.

In the last two decades, the Lorentz symmetry breaking has been intensively discussed in the literature (for a review, see for instance \cite{Bietenholz}), since it was discovered that there are interactions in the String Theory  which can lead to spontaneous Lorentz symmetry  breaking \cite{Colladay:1996iz}. Any violation of the Lorentz symmetry implies in a violation of the CPT symmetry, and, in general, the Lorentz-violating Standard Model Extension (SME) presents both CPT-even and CPT-odd Lorentz-breaking terms. 

Quantum electrodynamics (QED) is the theory with best agreement with experiments that we have, therefore it is an important source of investigation in the search for violations of the Lorentz symmetry. Several aspects of Lorentz-violating QED have been studied through the last decade (see f.e. \cite{Kostelecky:2001jc,Barraz:2007mi, Casana:2008nw, Casana:2009dq, Kostelecky:2009zp, Casana:2009xf, Ferrero:2009jb, Casana:2010nd, Gomes:2009wn, Casana:2011vh, Agostini:2011qk}), imposing bounds over Lorentz-violating parameters present in the SME Lagrangian~\cite{Kostelecky:2002hh, Kostelecky:2006ta, Kostelecky:2008ts,Charneski:2012py, Gomes:2014kaa}.

On the other hand, taking SUSY to be a fundamental symmetry of nature, it is important to ask whether it is possible to define a supersymmetric Lorentz-breaking field theory. One possible approach to this problem was presented by Belich et.al.~\cite{Hel}, where the authors included extra fields depending on the Lorentz-breaking parameters, allowing the arising of the CPT-odd Lorentz-breaking terms  in the supersymmetric Lagrangian. Another possibility is based on the Kostelecky-Berger construction \cite{KostBer} in which the main characteristic is the deformation of the SUSY algebra, allowing the arising of the CPT-even Lorentz-breaking terms. Following this last idea, the aether-superspace was constructed for both three and four-dimensional spacetime~\cite{Farias:2012ed,Lehum:2013pca}.

This work is devoted to construct the two-dimensional aether-superspace and to study the dynamical generation of mass in a Lorentz violating supersymmetric  QED in two dimensions of spacetime, $D=(1+1)$. QED in the two-dimensional spacetime is a very important theoretical laboratory because it is a massive photon gauge-invariant theory and it is an example of a  confining field theory~\cite{Schwinger:1962tp}. Using the aether-superspace techniques, we showed that the model presents a consistent dynamical generation of mass to the gauge aether-superfield and we observed that the deformed supersymmetry is not broken through radiative corrections. We also discuss some features to the gauge superfield propagation and its dispersion relation.     

%The paper is organized as follows. In Sec. \ref{2daether}, we present the two-dimensional aether superspace and define the action of an Abelian gauge superfield theory with a general superpotential. We also present its physical content writing the action of the model in terms of the component fields. In Sec. \ref{sqed2}, we study the propagation of the gauge superfield in the supersymmetric quantum electrodynamics in the two-dimensional aether-superspace showing that the model exhibits  dynamical generation of mass. We also discuss some features to the gauge superfield propagation and its dispersion relation. In Sec. \ref{remarks} we present our final remarks.

\section{Two-dimensional aether superspace} \label{2daether}

The usual conventions and notations of three-dimensional superspace can be directly applied to the supersymmetric field theories defined in the two-dimensional spacetime~\cite{gps}. Therefore, we will use the conventions and notations as close as possible to Ref.\cite{SGRS}. Just as the three and four-dimensional cases \cite{Farias:2012ed,Lehum:2013pca}, the deformation of the usual two-dimensional superspace to the two-dimensional aether-superspace is stated deforming the usual SUSY generators $Q_{\alpha}=i\left(\partial_{\alpha}-i{\theta}^{{\beta}}\gamma^m_{\beta\alpha}\partial_m\right)$  as 
\begin{eqnarray}
\mathcal{Q}_{\alpha}&=&i\left(\partial_{\alpha}-i{\theta}^{{\beta}}\gamma^m_{\beta\alpha}\tilde\partial_m\right),
\end{eqnarray}
\noindent where $\tilde\partial_m=\partial_m+k_{mn}\partial^n$, with the deformed operators $\mathcal{Q}$ satisfying the anti-commutation relation
\begin{eqnarray}
\{\mathcal{Q}_{\alpha},\mathcal{Q}_{\alpha}\}=2i\gamma^m_{{\alpha}\beta}\tilde\partial_m.
\end{eqnarray}
\noindent The supercovariant derivatives which anti-commute with $\mathcal{Q}_\alpha$ is defined by
\begin{eqnarray}\label{sder3d}
\tilde{D}_{\alpha}=\partial_{\alpha}+i{\theta}^{{\beta}}\gamma^m_{{\beta}\alpha}\tilde\partial_m~,
\end{eqnarray}
\noindent where Greek letters represent spinorial indices and Latin indices assume values of the two-dimensional spacetime coordinates ($0,1$). $\partial_{\alpha}$ is the derivative with respect to
the Grassmannian coordinates $\theta^{\alpha}$ and $k_{mn}$ is a constant tensor assuming an aether-like form $k_{mn}=\alpha u_mu_n$. The vector $u^m$ is a constant vector with $u^mu_m$ being equal either to $1$, $-1$ or $0$, and $\alpha$ is small \cite{aether,aether1}.

Such operators act on functions defined in this deformed superspace. Just as usual superspace, these functions (superfields) are terminating Taylor series in $\theta$, where the scalar aether-superfield $\Phi(x,\theta)$ can be decomposed as
\begin{eqnarray}\label{compon1}
\varphi=\Phi\Big{|}_{\theta=0}~,~~\psi^\alpha=\tilde{D}^{\alpha}\Phi
\Big{|}_{\theta=0}~,~~F=\tilde{D}^2\Phi\Big{|}_{\theta=0}~,
\end{eqnarray}
\noindent and the spinor aether-superfield $\Gamma_\alpha(x,\theta)$ as 
\begin{eqnarray}\label{compon2}
\chi_\alpha=\Gamma_\alpha\Big{|}_{\theta=0}~&,&~~B=\frac{1}{2}\tilde{D}^{\alpha}
\Gamma_\alpha\Big{|}_{\theta=0}~,\nonumber\\
V_{\alpha\beta}=-\frac{i}{2}\tilde{D}_{(\alpha}\Gamma_{\beta)}
\Big{|}_{\theta=0}~&,&~~\lambda_\alpha=\frac{1}{2}\tilde{D}^{\beta}\tilde{D}_{\alpha}\Gamma_\beta
\Big{|}_{\theta=0}~,
\end{eqnarray}
\noindent where $V_{\alpha\beta}=(\gamma^m)_{\alpha\beta}A_m$.

We can construct a supergauge field theory in this deformed superspace defining an action such as
\begin{eqnarray}\label{action3db}
S&=&-\frac{1}{2}\int d^4z\left[\frac{1}{2}W^\alpha W_\alpha+(\overline{\tilde{\nabla}^{\alpha}{\Phi}})
(\tilde{\nabla}_{\alpha}{\Phi})-f(\bar{\Phi}\Phi)\right]\nonumber\\
&=&-\frac{1}{2}\int d^4z \Big[\frac{1}{2}W^\alpha W_\alpha+(\overline{\tilde{D}^{\alpha}{\Phi}})(\tilde{D}_{\alpha}{\Phi})-i
\overline{\tilde{D}^{\alpha}{\Phi}}\Gamma_\alpha\Phi\nonumber\\
&&~~~~~~~~~~~~~~+i\Gamma^\alpha\bar{\Phi}
\tilde{D}_{\alpha}{\Phi}
  +\Gamma^{\alpha}\Gamma_{\alpha}\bar{\Phi}\Phi- f(\bar{\Phi}\Phi)\Big],
\end{eqnarray}
\noindent where the gauge covariant aether-superfield strength is defined as $W_\alpha=\dfrac{1}{2}\tilde{D}^{\beta}\tilde{D}_{\alpha}\Gamma_{\beta}$ and $f(\bar{\Phi}\Phi)$ is a general superpotential constructed by some power of the bilinear $\bar\Phi\Phi$.  The supergauge transformation of the matter aether-superfield $\Phi$ is $\Phi\rightarrow\mathrm{e}^{iK}\Phi$, and the spinor gauge aether-superfield transforms as
$\Gamma_{\alpha}'=\Gamma_{\alpha}+\tilde{D}_\alpha K$, where $K=K(z)$ is a real scalar aether-superfield. Notice that the supergauge transformations themselves are defined as deformed ones. The superspace volume element $d^4z$ stands for $d^2xd^2\theta$.

We can write the above action in terms of the components of the aether-superfields, Eqs. (\ref{compon1}) and (\ref{compon2}), revealing its physical content. Integrating over $d^2\theta$, Eq.(\ref{action3db})  can be cast as
\begin{eqnarray}
S&=&\int{d^2x}~\Big{\{}\lambda^\alpha i{(\gamma^m)_{\alpha}}^{\beta}\tilde\partial_m\lambda_\beta-\frac{1}{4}(\tilde\partial_mA_n-\tilde\partial_nA_m)^2\nonumber\\
&&+\bar{F}F+\bar{\psi}^\alpha{(\gamma^m)_{\alpha}}^{\beta}
[i\tilde\partial_m-A_m]\psi_{\beta} 
+(i\bar{\psi}^{\alpha}\lambda_\alpha\varphi+h.c.)\nonumber\\
&&+(\tilde\partial^m-iA^m)\bar{\varphi}(\tilde\partial_m+iA_m)\varphi+\frac{1}{2}
f'(\bar{\varphi}\varphi)[\bar{F}\varphi+\bar{\varphi}F+2\bar{\psi}^\beta
\psi_\beta]\nonumber\\
&&+\frac{1}{2}f''(\bar{\varphi}\varphi)[2\bar{\varphi}\varphi\bar{\psi}^\beta
\psi_\beta+\varphi^2\bar{\psi}^\beta\bar{\psi}_\beta+\bar{\varphi}^2\psi^\beta
\psi_\beta]
\Big{\}}, \label{action3dbb}
\end{eqnarray}

\noindent where $f'(\bar{\varphi}\varphi)=
\dfrac{\partial f(\bar{\Phi}\Phi)}{\partial(\bar{\Phi}\Phi)}
\Big{|}_{\bar{\Phi}\Phi=\bar{\varphi}\varphi}$, $f''(\bar{\varphi}\varphi)=
\dfrac{\partial^2 f(\bar{\Phi}\Phi)}{\partial(\bar{\Phi}\Phi)^2}
\Big{|}_{\bar{\Phi}\Phi=\bar{\varphi}\varphi}$ and $\tilde\partial_m=\partial_m+k_{mn}\partial^n$.

The photon is represented by the (gauge-fixed) vector field $A^m$ and the electron by the field $\psi_\alpha$, while their superpartners photino and seletron are represented by $\lambda_\alpha$ and $\varphi$, respectively. $F$ is an auxiliary field which can be eliminated using its equation of motion. 

\section{Quantum electrodynamics in the two-dimensional aether-superspace}\label{sqed2}

In order to quantize the model, our starting point is to add to the classical action, Eq.(\ref{action3db}), a gauge-fixing and the corresponding Fadeev-Popov terms given by
\begin{eqnarray}\label{eq1}
S_{FP}&=&\int{d^2xd^2\theta}\Big{\{}-\frac{1}{4\xi} \tilde{D}^{\alpha}\Gamma_{\alpha}\tilde{D}^2\tilde{D}^{\beta}\Gamma_{\beta}
+\bar{c}\tilde{D}^{2}c\Big{\}}.
\end{eqnarray}
\noindent Since we are interested in studying the dynamical generation of mass in such model, we have set $f(\bar\Phi\Phi)=0$, corresponding to a massless supersymmetric QED. Notice that ghosts do not couple to the physical aether-superfields because we are working in an Abelian theory.    

With the help of Eq.(\ref{eq1}), we derive the propagators of the model as
\begin{eqnarray}\label{eq4}
\langle \Gamma^{\alpha}(-\tilde{p},\theta_1)\Gamma^{\beta}(\tilde{p},\theta_2)\rangle&=&\frac{i}{2}\frac{\tilde{D}^2}{(\tilde{p}^2)^2}
\left(\tilde{D}_{\beta}\tilde{D}_{\alpha}-\xi \tilde{D}_{\alpha}\tilde{D}_{\beta}\right)\delta_{12}~,\\
\langle c(\tilde{p},\theta_1)\bar{c}(-\tilde{p},\theta_2)\rangle&=&i\frac{\tilde{D}^2}{\tilde{p}^2}\delta_{12}~,\\
\langle \Phi(\tilde{p},\theta_1)\bar\Phi(-\tilde{p},\theta_2)\rangle&=&-i\frac{\tilde{D}^2}{\tilde{p}^2}\delta_{12},
\end{eqnarray}

\noindent where $\delta_{12}=\delta^2(\theta_1-\theta_2)$ and $\tilde{p}_m=p_m+k_{mn}p^n$,
$\tilde{p}^2=p^2+2k_{mn}p^mp^n+k^{mn}k_{ml}p_np^l$,
$\tilde{D}^2=\partial^2-\theta^\beta(\gamma^m)_{\beta\alpha}\tilde{p}_m\partial^\alpha
+\theta^2\tilde{p}^2$. 

As mentioned above, this model is massless at classical level, but the electric charge $e$ is a dimensionful parameter with mass dimension one. Nonperturbative techniques can be used to see that QED in two-dimensional spacetime actually describes a free massive photon, because the electron and positron are confined~\cite{abdalla}. The dynamical generation of mass can also be computed perturbatively~\cite{ZinnJustin:2002ru}, resulting in the same conclusion. 

It is important to remark that a gauge-invariant mass can be also generated in a three-dimensional spacetime~\cite{jackiw}. But in three dimensions the gauge-invariant mass appears due to the generation of a topological Chern-Simons term, in models containing parity-violating interactions. In two dimensional spacetime the generation of mass is similar to the Higgs mechanism, due to the confinement of the electron and positron~\cite{abdalla,ZinnJustin:2002ru}.  In particular, through superfield techniques, the supersymmetric QED was shown to be finite to all loop orders in two~\cite{Gomes:2011aa} and three~\cite{lehum} dimensions of spacetime.
  
Let us compute the one-loop correction to the self-energy process of the gauge aether-superfield to show the perturbative dynamical generation of mass in the aether-superspace. Such contribution can be cast as
\begin{eqnarray}\label{eq8}
S_{1l}&=&-\frac{e^2}{2}\int{\frac{d^2p}{(2\pi)^2}}d^2\theta\int{\frac{d^2q}{(2\pi)^2}}\Gamma^{\alpha}(\tilde{p},\theta)\Big{\{}\frac{C_{\beta\alpha}}{\tilde{q}^2}-\frac{C_{\beta\alpha}}{(\tilde{q}+\tilde{p})^2}+\frac{1}{4}
\frac{\left(\tilde{p}^2C_{\beta\alpha}+\tilde{p}_{\beta\alpha}\tilde{D}^2\right)}{(\tilde{q}+\tilde{p})^2\tilde{q}^2}\Big{\}}\Gamma^{\beta}(-\tilde{p},\theta).\nonumber
\end{eqnarray}

The logarithmic divergences cancel each other, first and second terms between the curly brackets, so the one-loop contribution to the quadratic part of effective action for the gauge aether-superfield is
\begin{eqnarray}\label{eq8a}
S_{1l}&=&-\frac{e^2}{8}\int{\frac{d^2p}{(2\pi)^2}}d^2\theta~\Gamma^{\alpha}(\tilde{p},\theta)\Big{\{}\int{\frac{d^2q}{(2\pi)^2}}\frac{\left(\tilde{p}^2~C_{\beta\alpha}+\tilde{p}_{\beta\alpha}\tilde{D}^2\right)}{(\tilde{q}+\tilde{p})^2\tilde{q}^2}\Big{\}}\Gamma^{\beta}(-\tilde{p},\theta).
\end{eqnarray}

Summing up the tree level,
\begin{eqnarray}\label{eq1a} S_{c}&=&\int{\frac{d^2p}{(2\pi)^2}}d^2\theta\Big{\{}-\frac{1}{4}\Gamma_{\gamma}(\tilde{p},\theta)\tilde{p}^2\left(C_{\beta\gamma}+\frac{\tilde{p}_{\beta\gamma}\tilde{D}^2}{\tilde{p}^2}\right)\Gamma_{\beta}(-\tilde{p},\theta)\Big{\}}, 
\end{eqnarray}
\noindent and one-loop contribution given by Eq.(\ref{eq8a}), the effective action can be cast as
\begin{eqnarray}\label{eq8b}
S_{eff}&=&-\frac{1}{4}\int{\frac{d^2p}{(2\pi)^2}}d^2\theta~\Gamma^{\alpha}(\tilde{p},\theta)~\tilde{p}^2\left(C_{\beta\alpha}+\frac{\tilde{p}_{\beta\alpha}\tilde{D}^2}{\tilde{p}^2}\right)\Big[1+
\int{\frac{d^2q}{(2\pi)^2}}\frac{e^2/2}{\tilde{q}^2(\tilde{q}+\tilde{p})^2}\Big]\Gamma^{\beta}(-\tilde{p},\theta)\nonumber\\
&=&-\frac{1}{4}\int{\frac{d^2p}{(2\pi)^2}}d^2\theta~\Gamma^{\alpha}(\tilde{p},\theta)~\left(C_{\beta\alpha}+\frac{\tilde{p}_{\beta\alpha}\tilde{D}^2}{\tilde{p}^2}\right)\tilde{p}^2\left(1+\frac{e^2 \Delta}{8\pi~\tilde{p}^2}\right)\Gamma^{\beta}(-\tilde{p},\theta)\nonumber\\
&=&-\frac{1}{4}\int{\frac{d^2p}{(2\pi)^2}}d^2\theta~\Gamma^{\alpha}(\tilde{p},\theta)~\left(C_{\beta\alpha}+\frac{\tilde{p}_{\beta\alpha}\tilde{D}^2}{\tilde{p}^2}\right)\left(\tilde{p}^2+\frac{e^2 \Delta}{8\pi}\right)\Gamma^{\beta}(-\tilde{p},\theta),
\end{eqnarray}

\noindent where the integral could be evaluated by a changing of variable $q$ to $\tilde{q}$, where we can write $\int{d^2q}=\Delta\int{d^2\tilde{q}}$, with $\Delta=\det{(\frac{\partial
    q^m}{\partial\tilde{q}^n})}=\det^{-1}(\delta^m_n+k^m_n)=(1+\alpha u^mu_m)^{-1}$ being the Jacobian of the changing of variable. Notice we find the presence of a massive pole for the perturbative full propagator, $M^2= e^2\Delta/(8\pi^2)$ (that is dependent on the aether properties through $\Delta$), since in two-dimensional spacetime the electric charge $e$ has mass dimension one. The massive pole is the same for the superpartners $A^m$ and $\lambda_\alpha$ indicating that the deformed supersymmetry is manifest.

Integrating over the Grassmannian variables, Eq.(\ref{eq8b}) can be written as 
\begin{eqnarray}\label{action2dcomp}
S&=&\int{\frac{d^2p}{(2\pi)^2}}~\Big{\{}-\frac{1}{2}A^m(\tilde{p})\left(\eta_{mn}-\frac{\tilde{p}_m\tilde{p}_n}{\tilde{p}^2}\right)\left(\tilde{p}^2+\frac{e^2\Delta}{8\pi}\right)A^n(-\tilde{p})\nonumber\\
&& \hspace{2cm}+\lambda^\alpha(\tilde{p}) {(\gamma^m)_{\alpha}}^{\beta}\tilde{p}_m\left(\tilde{p}^2+\frac{e^2\Delta}{8\pi}\right)\lambda_\beta(-\tilde{p})
\Big{\}}.
\end{eqnarray}

Notice that the gauge aether-superfield satisfies the dispersion relation of a massive particle $\tilde{p}^2+M^2=p^2+2k_{mn}p^mp^n+k^{mn}k_{ml}p_np^l+M^2$, which has the structure similar for the propagators in the CPT-even Lorentz-breaking models (see e.g.~\cite{Casana:2010nd,Fer1,Fer3}).

\section{Final remarks}\label{remarks}

The massless quantum electrodynamics in a two-dimensional spacetime is a well-known model which presents dynamical generation of mass and it is an example of confining field theory. Even though this theory classically describes a massless fermion in electromagnetic interaction, its quantum description reveals a non interacting massive gauge field theory. The dynamical generation of mass can be evaluated perturbatively and, as discussed by Schwinger~\cite{Schwinger:1962tp}, it is an exact result to all orders.

The supersymmetric (Lorentz-violating) version of this model presented in this paper has two additional degrees of freedom, that are the superpartners of the electron and photon: the scalar field $\varphi$ (seletron) and the fermionic field $\lambda_\alpha$ (photino), respectively. We observed that the dynamically generated masses are the same to the photon and its superpartner photino, implying that the deformed supersymmetry is not broken. Moreover, the genarated mass is dependent on the aether properties. We have also showed that the dispersion relation of the (super) photon is similar to that observed in CPT-even Lorentz-breaking models.  To do this, we have computed the one-loop correction to the propagation of the aether-superfield using techniques of two-dimensional aether-superspace.

The techniques developed here can easily be applied to the study of non-Abelian gauge theories, in particular in the study of Lorentz violations in a two-dimensional supersymmetric quantum chromodynamics.

%\vspace{1cm}
\acknowledgments{ This work was partially supported by Conselho Nacional de
Desenvolvimento Cient\'{\i}fico e Tecnol\'{o}gico (CNPq) and Funda\c{c}\~{a}o de Apoio \`{a} Pesquisa do Rio Grande do Norte (FAPERN). The author would like to thank L. Ibiapina Bevil\'aqua for the careful reading.}

\end{document}